\documentclass[aps,prl,twocolumn,showpacs]{revtex4-1}
\usepackage{graphicx,amsmath,amssymb}
\usepackage[usenames]{color}
\begin{document}

\title{
From quantum Rabi model to Jaynes-Cummings model: symmetry-breaking quantum phase transitions, topological phase transitions and multicriticalities
}

\author{Zu-Jian Ying }
\email{yingzj@lzu.edu.cn}
\affiliation{School of Physical Science and Technology, Lanzhou University, Lanzhou 730000, China}

\begin{abstract}
We study the ground state (GS) and excitation gap of anisotropic quantum Rabi model (QRM) which connects the fundamental QRM and the Jaynes-Cummings model (JCM). While the GS has a second-order quantum phase transition (QPT) in the low frequency limit, turning on finite frequencies we shed a novel light on the phase diagram to illuminate a fine structure of first-order transition series. We find the QPT is accompanied with a hidden symmetry breaking, whereas the emerging series transitions are topological transitions without symmetry breaking. The topological structure of the wave function provides a novel universality classification in bridging the QRM and the JCM. We show that the conventionally established triple point is actually a quintuple or sextuple point and following the penta-/hexa-criticality emerge a series of tetra-criticalities.
\end{abstract}
\pacs{ }
\maketitle


\textit{Introduction.}--In the past decade the frontiers of quantum physics
have witnessed both experimental \cite{Diaz2019RevModPhy,Kockum2019NRP}
and theoretical \cite{Braak2011,Boite2020} advances in the study of the
light-matter interaction. The experimental access to ultrastrong coupling
regime\cite{Diaz2019RevModPhy,Wallraff2004,Gunter2009,Niemczyk2010,Peropadre2010,FornDiaz2017, Forn-Diaz2010,Scalari2012,Xiang2013,Yoshihara2017NatPhys,Kockum2017} has brought a rich phenomenology unexpected in 
weak couplings \cite{Diaz2019RevModPhy,Kockum2019NRP}. One the other hand, the recent milestone
theoretical work of revealing the Braak integrability \cite{Braak2011} has
attracted tremendous attention on the quantum Rabi model (QRM) and its extensions\cite%
{Boite2020,Wolf2012,FelicettiPRL2020,Felicetti2018-mixed-TPP-SPP,Felicetti2015-TwoPhotonProcess,Simone2018,
Solano2011,ChenQH2012,e-collpase-Duan-2016,ZhengHang2017,Irish2014,PRX-Xie-Anistropy,Batchelor2015,XieQ-2017JPA,Hwang2015PRL,Bera2014Polaron,
Ying2015,LiuM2017PRL,CongLei2017,CongLei2019,Ying-2018-arxiv,Ying2020-nonlinear-bias,PengJie2019,Liu2015,Ashhab2013,
ChenGang2012,FengMang2013,ZhangYY2016,ChenGang2012,
AnistropicShen2017,Eckle-2017JPA,Casanova2018npj}.
The QRM \cite{rabi1936} and the Jaynes-Cummings model (JCM) \cite{JC-model} are the most
fundamental models for the light-matter interaction. They are also
fundamental building blocks for quantum information and quantum computation
\cite{Diaz2019RevModPhy,Romero2012} and closely connected to models in
condense matter \cite{Kockum2019NRP}.

The essential difference of the QRM and\ the JCM lies in the
counter-rotating interaction. Both experimental measurements \cite%
{Forn-Diaz2010,Pietikainen2017,Yimin2018} and theoretical studies\cite%
{PRX-Xie-Anistropy,LiuM2017PRL} have raised the concern on the role of the
counter-rotating interaction, a full understanding of which is one of the
central topics in investigations of light-matter interaction. The coupling
anisotropy tunes the strength of the counter-rotating interaction, thus the
anisotropic QRM\ is the bridge of the two fundamental models. The
anisotropic QRM is integrable\cite{PRX-Xie-Anistropy} and interestingly the
model exhibits a few-body quantum phase transition which can be connected to
the many-body and thermodynamical cases via the scaling of critical
components\cite{LiuM2017PRL}.

Generally the study of phase transitions has stimulated the developments of
theoretical physics in deepening the understanding of nature. The
traditional Landau theory\cite{Landau-theory} made a breakthrough to realize
that different phases correspond to the realization of different symmetries
and a phase transition undergoes a symmetry breaking. In the last decades, a
new class of phases was discovered coined the term "topological phases"\cite%
{Topo-KT-transition,Topo-KT-NoSymBreak,Topo-Haldane-1,Topo-Haldane-2,Topo-Wen}%
. The occurrence of topological phase transition requires gap closing but
without need of any symmetry breaking. Traditionally these phase transitions
occur in macroscopic systems in condensed matter. Recently few-body quantum
phase transition has received special attention \cite%
{Hwang2015PRL,LiuM2017PRL,Ashhab2013,Ying2015,Ying-2018-arxiv,Zhu2020PRL}. Various patterns of
symmetry breaking can occur to induce tricriticalities and quadruple points
even in a single-qubit system\cite{Ying2020-nonlinear-bias}. One may wonder
whether there is any analog of topological phase transition in few-body
systems.

In this work we study the anisotropic QRM to renew the understanding of the
role of the counter-rotating term. Rather than conventional consideration of
transitions in the low frequency limit, we shed light on the phase diagrams
at finite frequencies, which enables us to illuminate an emerging structure
of phase boundary series with gap closing. We find these emerging additional
phase transitions are actually topological transitions without symmetry
breaking. The number of zero points in the wave function turns out to be the
topological number in the novel classification of the phases. We also find
emerging series of multicriticalities beyond the conventional picture.

\textit{Model and symmetry.}--The anisotropic QRM\cite{PRX-Xie-Anistropy,LiuM2017PRL} reads
\[
H=\omega a^{\dagger }a+\frac{\Omega }{2}\sigma _{x}+g\left[ \left(
\widetilde{\sigma }_{-}a^{\dagger }+\widetilde{\sigma }_{+}a\right) +\lambda
\left( \widetilde{\sigma }_{+}a^{\dagger }+\widetilde{\sigma }_{-}a\right) %
\right]
\]%
where $\sigma _{x,y,z}$ is the Pauli matrix, $a^{\dagger }(a)$ creates
(annihilates) a bosonic mode with frequency $\omega $. The coupling strength
is denoted by $g$ and the anisotropy by $\lambda $. The QRM is retrieved by $%
\lambda =1$ and the JCM by $\lambda =0.$ By the transforms $a^{\dagger }=(%
\hat{x}-i\hat{p})/\sqrt{2},$ $a=(\hat{x}+i\hat{p})/\sqrt{2},$ $\hat{p}=-i%
\frac{\partial }{\partial x}$ and $\widetilde{\sigma }^{\pm }=(\sigma
_{z}\mp i\sigma _{y})/2,$ $\sigma _{x}=\sigma ^{+}+\sigma ^{-},$ $\sigma
_{y}=-i(\sigma _{+}-\sigma _{-})$ we can map to the effective spatial space
\begin{equation}
H=\frac{\omega }{2}\hat{p}^{2}+v_{\sigma _{z}}+[\frac{\Omega }{2}-g_{y}i%
\sqrt{2}\hat{p}]\sigma ^{+}+[\frac{\Omega }{2}+g_{y}i\sqrt{2}\hat{p}]\sigma
^{-}  \label{Hx}
\end{equation}%
where $g_{y}=\frac{\left( 1-\lambda \right) }{2}g,$ $g_{z}=\frac{\left(
1+\lambda \right) }{2}g$ and $\varepsilon _{0}^{y}=-\frac{1}{2}%
[g_{y}^{\prime 2}+1]\omega $. The harmonic potentials $v_{\sigma
_{z}}=\omega \left( x+g_{z}^{\prime }\sigma _{z}\right) ^{2}/2+\varepsilon
_{0}^{y}$ in the two spin components are shifted in opposite directions by
the coupling via $g_{z}^{\prime }=\sqrt{2}g_{y}/\omega $. Then the $\Omega $
term plays the role of spin flipping in spin $\sigma_z=\pm$ space and tunneling in the
effective spatial space \cite{Ying2015,Irish2014}. The $g_{y}$ term is
Rashba spin-orbit coupling (RSOC) in competition with the $\Omega $ term. The model at
any anisotropy possesses the parity symmetry $\hat{P}=\sigma
_{x}(-1)^{a^{\dagger }a}$ which commutes with $H$ \cite{PRX-Xie-Anistropy}.

\begin{figure}[t]
\includegraphics[width=1.0\columnwidth]{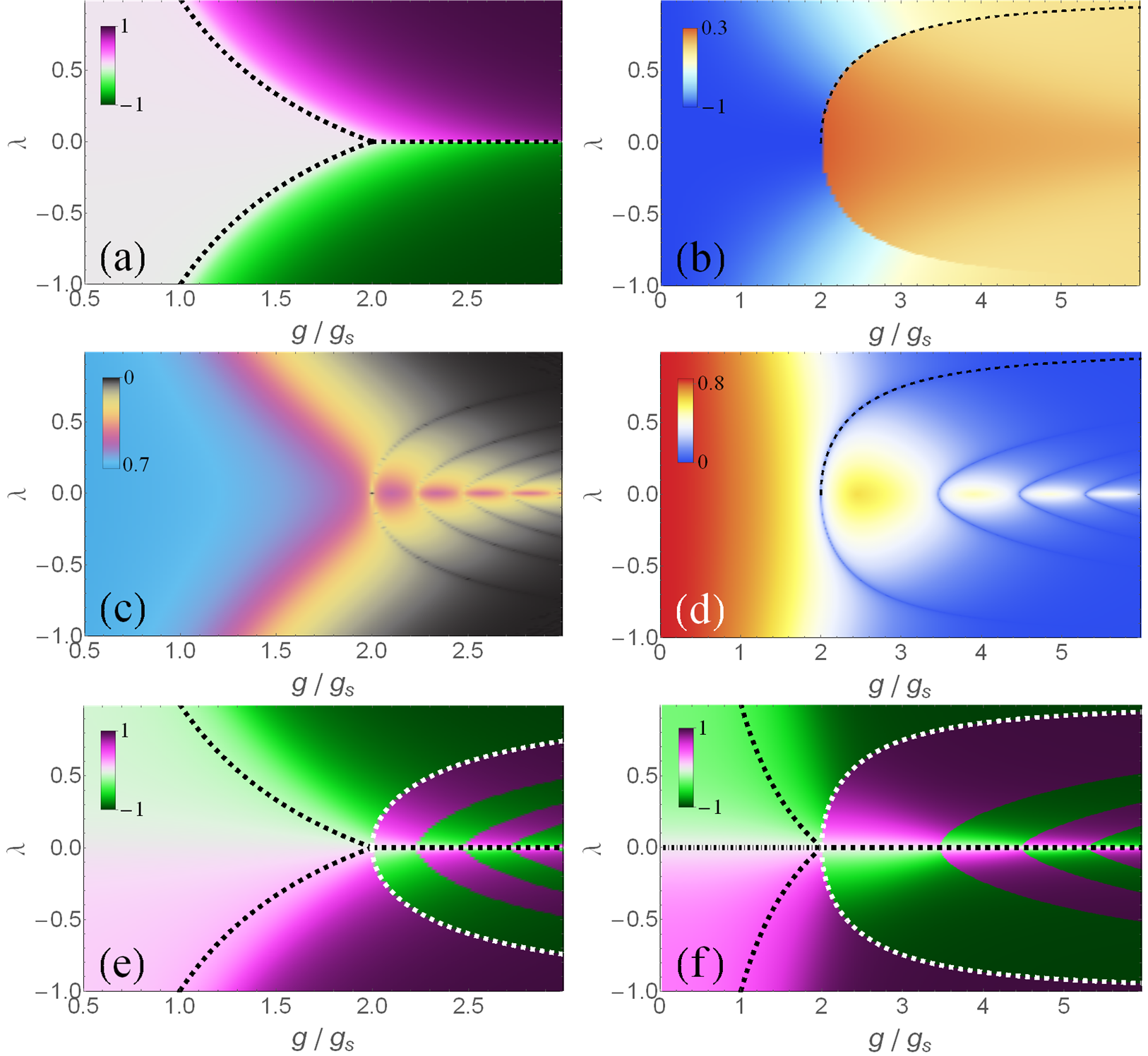}
\caption{(color online) \textit{Multicriticality in phase diagrams of
anisotropic QRM.} (a) $A=\langle a^{\dagger}a^{\dagger}\rangle/A_0$
of the ground state (GS) versus coupling $g$ and anisotropy $\protect\lambda$
at $\protect\omega =0.01\Omega$. $A_0=[(1+|\lambda|)g/(2\omega)]^2$ (b) $\langle \protect\sigma_x \rangle$ at
at $\protect\omega =2\Omega$. (c) First-excitation gap $\Delta$ at $\protect%
\omega =0.1\Omega$ (plotted by $\Delta ^{1/6}$ to show the boundaries
better). (d) $\Delta ^{1/4}$ at $\protect\omega =0.5\Omega$. (e) $A*P$ at $%
\protect\omega =0.1\Omega$. $P$ is parity. (f) $A*P$ at $\protect\omega %
=0.5\Omega$. The dashed lines in (a,b,d,e,d) are analytic boundaries.}
\label{fig-PhaseDiagrams}
\end{figure}

\textit{Conventional phase diagram.--}The anisotropic QRM has a quantum
phase transition in the low frequency limit \cite{LiuM2017PRL} at the
transition boundary $g_{c}^{\lambda }=\frac{2}{1+\left\vert \lambda
\right\vert }g_{\mathrm{s}}$ where $g_{\mathrm{s}}=\sqrt{\omega \Omega }/2$
is the transition point of the QRM. Above $g_{c}^{\lambda }$ the ground
states (GS) in positive and negative anisotropy are respectively $x$ and $p$
types, in the sense that the $\lambda >0$ regime has finite expectation $%
\langle \hat{x}^{2}\rangle $ but vanishing $\langle \hat{p}^{2}\rangle $
while it is reversed in $\lambda <0$ regime. So there are totally three
phases. We can distinguish the three phases conveniently by one physical
quantity $\langle a^{\dagger }a^{\dagger }\rangle $ which
is equivalent to the difference of $\langle \hat{x}^{2}\rangle $ and $%
\langle \hat{p}^{2}\rangle $, as illustrated in Fig.\ref{fig-PhaseDiagrams}%
(a) by exact diagonalization at $\omega =0.01\Omega $.

\textit{Excitation gap and novel phase diagram.-- }When we tune up the
frequency we illuminate a fine structure of the phase diagram beyond the
above conventional picture. In Fig.\ref{fig-PhaseDiagrams}(c) we show the
map of the first excitation gap at $\omega =0.1\Omega $, where one can see
that, besides the conventional boundaries, there are a series of additional
phase boundaries emerging. These boundaries show up in gap closing and
reopening. These additional boundaries become sparser when we raise the
frequency higher, as $\omega =0.5\Omega $ in panel (d). Although higher
boundaries are moving with the increase of frequency, the primary one of the
additional boundaries remains invariant. Let us label the primary one 
by $g_{\mathrm{T1}}$. It should be mentioned that, unlike the
second-order transition $g_{c}^{\lambda }$, the additional phase transitions
are of first order. This first order feature can be reflected in the spin
expectation $\langle \sigma _{x}\rangle $ as shown by Fig.\ref%
{fig-PhaseDiagrams}(b) with $\omega =2\Omega $, despite that the
discontinuity decreases with the coupling strength.

\textit{Multicriticality.--}The additional transition boundaries involves
changes of parity $P$. We notice that the parity is symmetric with respect
to\ JCM line at $\lambda =0$ while the $\langle a^{\dagger }a^{\dagger
}\rangle $ is antisymmetric. Combining the two quantities unveils a novel
phase diagram as in Fig.\ref{fig-PhaseDiagrams}(e) with $\omega =0.1\Omega $%
. We see that the conventional triple point becomes a quintuple point and
the tricriticality becomes a\ pentacriticality. Moreover, following the
pentacriticality a series of quadruple points and tetra-criticalities
emerge. Raising the frequency higher, as $\omega =0.5\Omega $ in Fig.\ref%
{fig-PhaseDiagrams}(f), the difference between the positive and negative $%
\lambda $ regimes also arises for the weak couplings below $g_{c}^{\lambda }$%
. This is coming from the breaking of the U(1) symmetry, i.e. the excitation
number $\hat{E}=a^{\dagger }a+\sigma _{x}$, which is a symmetry of
the JCM. Thus, if taking this U(1)-breaking boundary into account, the
novel pentacriticality would renew again to turn to be a hexacriticality.

\begin{figure}[t]
\includegraphics[width=0.88\columnwidth]{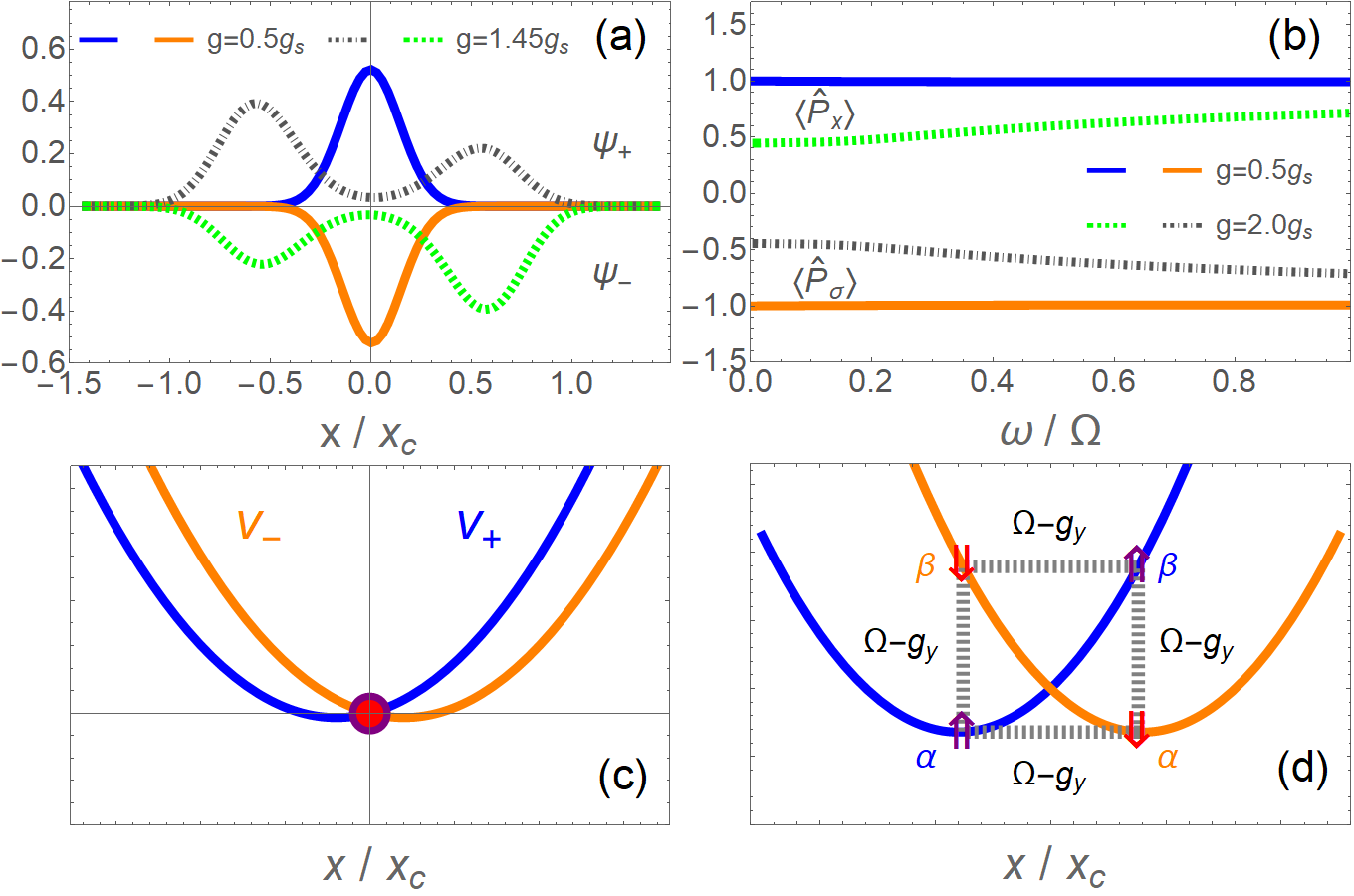}
\caption{(color online) \textit{Hidden symmetry breaking at the transition $%
g_c^\protect\lambda$. } (a) GS wave function $\protect\psi _{\pm}$ before
(solid lines, $g=0.5 g_\mathrm{s}$) and after transition (broken
lines, $g=1.45 g_\mathrm{s}$) at $\protect\lambda=0.5$ and $\protect\omega%
=0.01\Omega$. (b) $\langle \hat{P}_x\rangle$ and $\langle \hat{P}_\protect%
\sigma\rangle$ before and after transition. (c) Effective potential $v_ \pm $
before transition with dots marking the particle position. (d) $v_ \pm $
after transition. The arrows represent the spins and dashed lines label the
tunneling ($\Omega$) and RSOC coupling ($g_y$). $\protect\alpha,%
\protect\beta$ are wave-packet weights and $x_c=\sqrt{2}g_\mathrm{s}/\omega$. }
\label{fig-HiddenSymmetryBreaking}
\end{figure}

\textit{Hidden symmetry breaking at transition }$g_{c}^{\lambda }$\textit{.}%
--Before revealing the nature of the additional series transitions, let us
first re-visit the conventional transition $g_{c}^{\lambda }$ to unveil its
symmetry-breaking aspect. Although the anisotropic QRM possesses the parity
symmetry overall the parameter regime, we find the transition $%
g_{c}^{\lambda }$ also undergoes a symmetry breaking which is hidden. Let us
decompose the parity into two parts, $\hat{P}=\hat{P}_{\sigma }\hat{P}_{x}$
where $\hat{P}_{\sigma }=\sigma _{x}$ reverses the spin while$\ \hat{P}%
_{x}=(-1)^{a^{\dagger }a}$ effectively inverses the spatial space $%
x\rightarrow -x$ \cite{Ying2020-nonlinear-bias}. For the general cases of
the\ anisotropic QRM, both the spin reversion and the space inversion are
needed to achieve the parity symmetry, while generally the model does not
possess the individual symmetry of either $\hat{P}_{\sigma }$ or $\hat{P}%
_{x}$. However, the GS before the transition $g_{c}^{\lambda }$ is an
exception.

In fact, in this regime the GS is a Gaussian-like wave packet in each spin
component as shown by the solid lines in Fig.\ref{fig-HiddenSymmetryBreaking}%
(a). This is because, despite of the potential separation, in weak couplings
the effective particle tends to stay at the origin to gain a maximum spin
flipping or tunneling energy ($\Omega $ term), as sketched in Fig.\ref%
{fig-HiddenSymmetryBreaking}(c). The weak coupling $g$ provides a small
potential separation $g_{z}^{\prime }$. The RSOC coupling ($g_{y}$
term in Eq.\eqref{Hx}) not only depends on $g$ but also cancels itself for
the Gaussian wave packet at the origin. Thus the spin flipping is dominating
before the transition $g_{c}^{\lambda }$. The degenerate potential at the
origin gives an equal weight for the two spins, while a symmetric wave
packet yields a maximum wave-packet overlap favorable for gaining more spin
flipping energy. As a consequence, the state
possesses both $\hat{P}_{x}$ and $\hat{P}_{\sigma }$ symmetries. After the transition the wave function
splits into two packets, as shown in by the broken lines in Fig.\ref%
{fig-HiddenSymmetryBreaking}(a). As the packet weights on the two sides are
imbalanced due to the potential difference as indicated in Fig.\ref%
{fig-HiddenSymmetryBreaking}(d), neither symmetry of $\hat{P}_{x}$ and $\hat{P}_{\sigma }$ can be preserved. 
Therefore the conventional transition at $g_{c}^{\lambda }$ is accompanied with a hidden 
symmetry breaking of $\hat{P}_\sigma$ and $\hat{P}_x$. 
Fig.\ref{fig-HiddenSymmetryBreaking}(b) shows
the expectation of $\hat{P}_{\sigma }$ and $\hat{P}_{x}$ with amplitudes
remaining at $1$ before the transition indicating preserved symmetries and
deviation from $1$ after the transition meaning broken symmetries.

\begin{figure}[t]
\includegraphics[width=1.0\columnwidth]{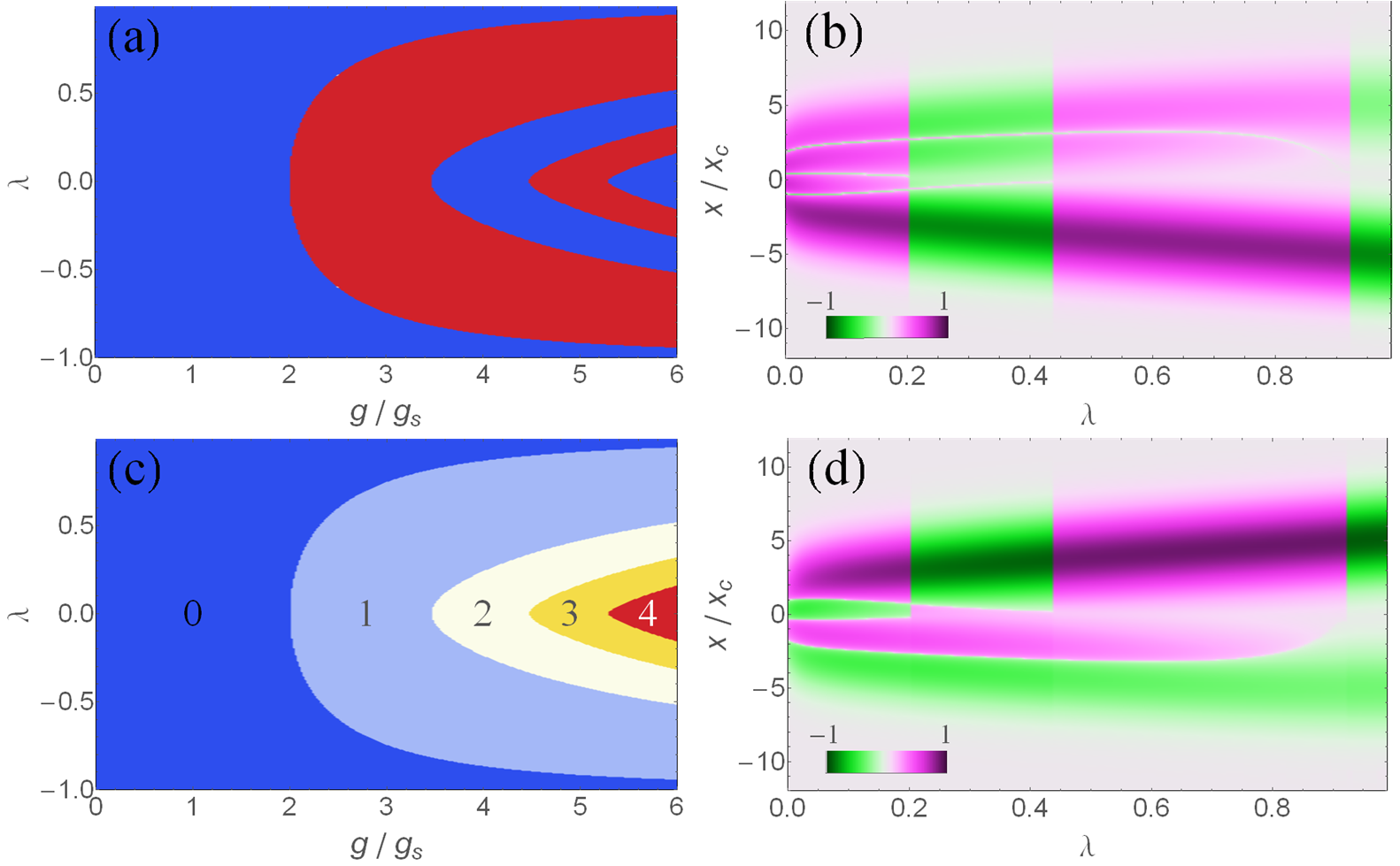}
\caption{(color online) \textit{Parity and topological transitions.} (a)
Phase diagram of the GS parity with $P=-1$ (blue) and $P=1$ (red). (b)
$\protect\psi _{+}(x)*\protect\psi _{-}(-x)$ versus $\protect%
\lambda$ at $g=5.2g_\mathrm{s}$. (c) Topological phase diagram. The numbers
mark zero numbers $n_\mathrm{Z}$. (d) $\protect\psi _{-}(x)$ versus $\protect%
\lambda$. Here $\protect\omega=0.5$ and the color contrast in (b,d) is
enhanced by amplified amplitude $|\protect\psi _{\pm}|^{1/4}$. }
\label{fig-parity-topology}
\end{figure}

\textit{Topological phase transitions.}--So far we have seen that the
transitions at $g_{c}^{\lambda }$ and at $\lambda =0$ are all accompanied
with some symmetry breaking. Now we look at the additional series
transitions. Since the parity symmetry is valid for all parameter regimes,
these additional transitions do not break the parity symmetry either. Indeed
the parity changes the sign at these transitions, as shown by Fig.\ref%
{fig-parity-topology}(a). However, the parity has only two values, not
enough to label the series phases separated by the additional transition
boundaries. In fact, we find that the additional transitions are topological
phase transitions which are different class of phase transition from those
transitions at $g_{c}^{\lambda }$ and at $\lambda =0.$ The phases separated
by these topological transitions are distinguished from each other by their
topological feature, each of them having an individual topological number.

Indeed, the wave function in each phase belt has a same number of zero
points $n_{\mathrm{Z}}$. In Fig.\ref{fig-parity-topology}(b), we show the
wave function by the product of the spin-up and spin-down components, $\psi
_{P}\left( x\right) =\psi _{+}\left( x\right) \ast \psi _{-}\left( -x\right)
$, in the variation of the anisotropy at a fixed coupling. Starting from the
QRM side at $\lambda =1$ we have $\psi _{P}\left( x\right) $ all
non-negative (yellow) in $x$ space, indicating the negative parity. Reducing
the value of $\lambda $ \ till the JCM side at $\lambda =0$, we have three
topological transitions. At each transition $\psi _{P}\left( x\right) $
changes between non-negative (yellow) and non-positive (blue), indicating a
changeover between negative and positive parities. A closer look at the
evolution of $\psi _{P}\left( x\right) $ in Fig.\ref{fig-parity-topology}%
(b), one can notice some white lines appearing. These white lines are
actually zero-value lines. Each phase has a same $n_{\mathrm{Z}}$ number,
thus forming lines. From the QRM to the JCM we have zero line number $n_{%
\mathrm{Z}}$=$0$, $1$, $2$, $3$ corresponding to the four phases. The zero
lines can be more clearly visualized from $\psi _{-}\left( x\right) $ as in
Fig.\ref{fig-parity-topology}(d), where the color shifts between yellow and
blue if tracking along the $x$ dimension. Given a fixed number $n_{\mathrm{Z}%
}$, one cannot go to another $n_{\mathrm{Z}}$ state by continuous
shape deformation of the wave function. Thus the zero number $n_{\mathrm{Z}}$
is the topological number of each phase. We extract the topological phase
diagram in Fig.\ref{fig-parity-topology}(c) where the numbers mark $n_{%
\mathrm{Z}}$. These topological phase transitions are accompanied with gap
closing, as in Fig.\ref{fig-PhaseDiagrams}(d). The afore-discussed
transition at $g_{c}^{\lambda }$ occurs within the topological phase $n_{%
\mathrm{Z}}=0$, without undergoing a topological transition in the
wave-packet splitting. The gap at $g_{c}^{\lambda }$ becomes small in
lowering frequency, but never exactly closes unless $\omega =0$.

\begin{figure}[t]
\includegraphics[width=1.0\columnwidth]{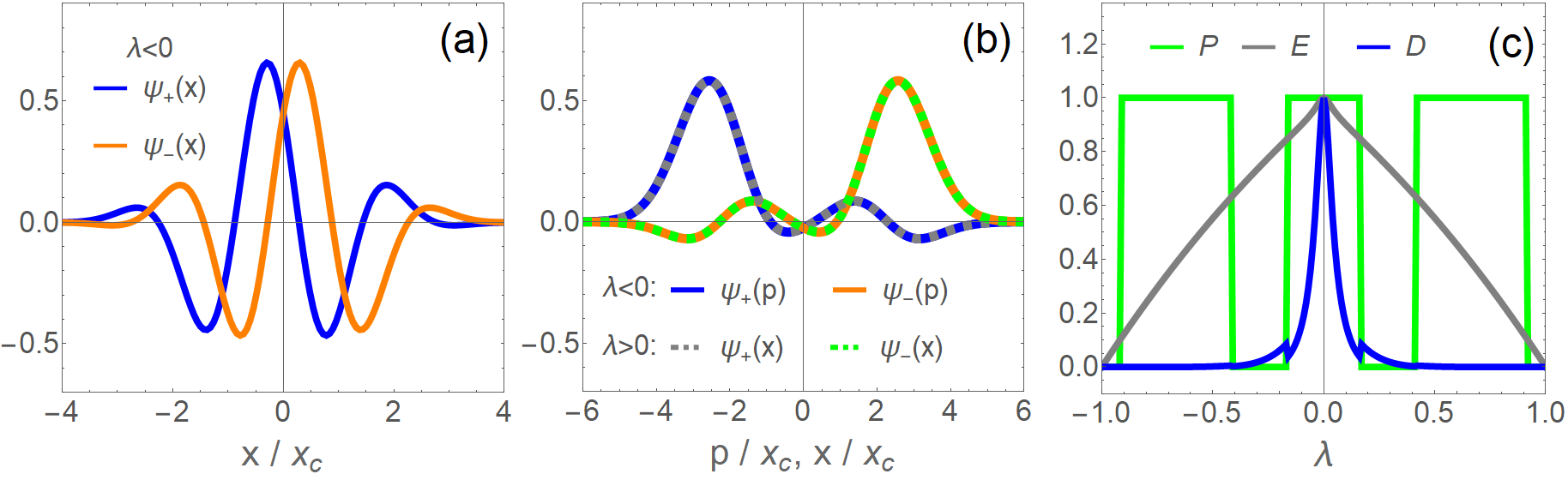}
\caption{(color online) \textit{$x$-$p$ duality and spontaneous symmetry
breaking.} (a) Wave function in $x$ space for $\protect\lambda = -0.1$ (b)
Wave function in $p$ space for $\protect\lambda =-0.1$ (solid) which
coincides with that of $\protect\lambda = 0.1$ (dashed) in $x$ space. (c)
Expectation of parity ($P=\langle \hat{P} \rangle$, plotted by $(P+1)/2$),
excitation number ($E = \langle \hat{n}+ \protect\sigma _x \rangle$, plotted
by normalized $E(1)-E(\protect\lambda)$ ) and $x$-$p$ duality ($D$). Here we
illustrate at $g=5g_ {\mathrm{s}}$ and $\protect\omega =0.5\Omega$. }
\label{fig-x-p-duality}
\end{figure}
\begin{figure}[t]
\includegraphics[width=0.9\columnwidth]{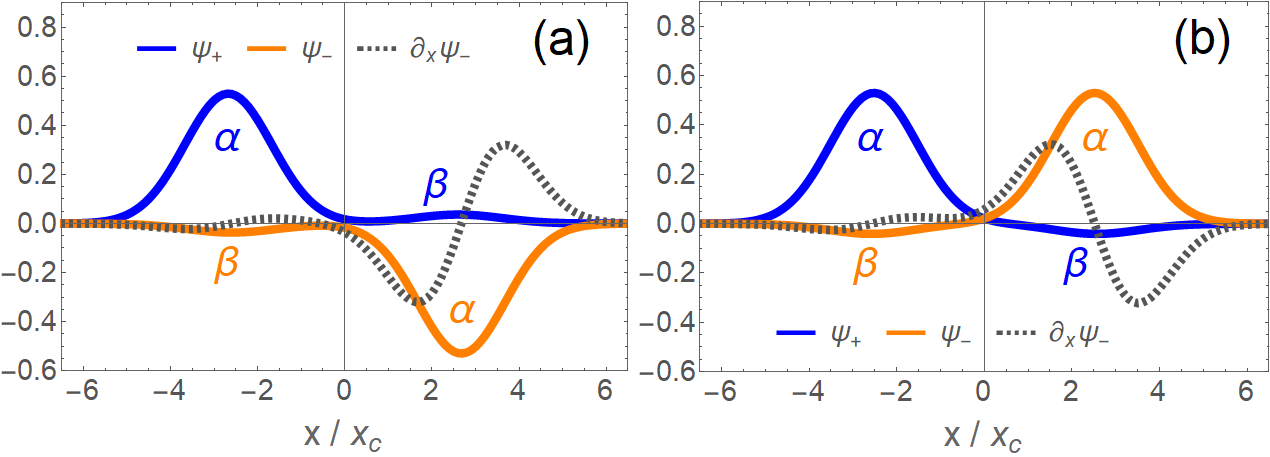}
\caption{(color online) \textit{Energy competition around the primary
topological transition.} $\protect\psi_{\pm}$ and $\partial _x\protect\psi%
_{\pm}$ before (a) and after (b) the transition $g_\mathrm{T1}$ at $\protect%
\lambda =0.7,0.8$ respectively. Here $g=3.0g_\mathrm{s}$ and $\protect\omega%
=0.5 \protect\omega$. }
\label{fig-mechanism}
\end{figure}

\textit{Negative }$\lambda $\textit{\ and }$x$\textit{-}$p$\textit{\ Duality}%
.--The afore-discussed topological transition in $x$ space refers to the $%
\lambda \geqslant 0$ regime. For the $\lambda <0$ regime, the quantum state
is $p$ type as mentioned in the conventional phase diagram. Thus
topological character should lie in the $p$ space instead of the $x$ space.
Indeed by the spin rotation and $x$-$p$ exchange $\{\sigma
_{x},\sigma _{y},\sigma _{z}\}\rightarrow \{\sigma _{x},-\sigma _{z},\sigma
_{y}\},\ -i\partial _{x}\rightarrow p,\ x\rightarrow i\partial _{p}$ one can
get the dual form of Hamiltonian \eqref{Hx}\textbf{\ }in the $p$ space, with
$g_{y}$ and $g_{z}$ exchange the roles as now $\lambda $ is negative. In Fig.%
\ref{fig-x-p-duality}(a) we illustrate the wave function in the $x$ space
for $\lambda <0$, which seemingly does not match the topological feature
extracted above. However as shown in Fig.\ref{fig-x-p-duality}(b), after the
mapping to the $p$ space, the wave function exactly becomes the same as that
in the $x$ space from $\lambda >0$ regime, up to an irrelevant total phase.

\textit{Spontaneous symmetry breaking (SSB).}---The JCM is invariant under the
above mapping as $g_{y}=g_{z}$ at $\lambda =0$, and it is also easy to show
analytically that the JCM\ wave function, which has a form $\psi _{n\pm
}^{\left( z\right) }=C_{n}\phi _{n}\mp C_{n}\phi _{n+1}$ where $\phi _{n}$
is the quantum Harmonic oscillator, is invariant under the
duality mapping. Let us denote such a duality symmetry by $D$. Thus the JCM
has the parity symmetry $P$, the U(1) symmetry (excitation number $E$) and
the duality symmetry $D.$ It is worth having a comparison of how these
symmetries are affected by the anisotropy. Fig.\ref{fig-x-p-duality}(c) we
plot the expectations of the symmetry operations versus $\lambda $. In
contrast to preserved $P$ and and mildly broken $E$, above $g_{c}^{\lambda }$
there is a SSB behavior in $D$, sharper for lower
frequencies and stronger couplings. Below $g_{c}^{\lambda }$ (unplotted),
the variation in $E$ remains similar to above $g_{c}^{\lambda }$, while the
sharp change of $D$ is flatten and $D$ even becomes unbroken in low
frequency limit. Under the protection of parity, topological transition is 
however unaffected by the $D$-SSB, but triggered by gap closing.

\textit{Wave-function braiding and gap closing.}--As afore-discussed, after
the transition $g_{c}^{\lambda }$ the GS wave function splits into two
packets, which can be represented by\cite{Ying2015} $\psi _{+}\left( x\right) =\alpha
\varphi _{\alpha }+\beta \varphi _{\beta }$ with weights $\alpha ,\beta $ and $\psi _{-}\left( x\right)
=-\psi _{+}\left( -x\right) $ under the parity symmetry. As sketched in Fig.\ref%
{fig-HiddenSymmetryBreaking}(d), there are four channels of tunneling and
RSOC, i.e. $\Omega _{\gamma \gamma ^{\prime }}=-\frac{%
\Omega }{2}\ \gamma \gamma ^{\prime }\langle \varphi _{\gamma }\left(
x\right) |\varphi _{\gamma ^{\prime }}\left( -x\right) \rangle $ and $%
g_{\gamma \gamma ^{\prime }}^{y}=\sqrt{2}g_{y}\gamma \gamma ^{\prime
}\langle \varphi _{\gamma }\left( x\right) |\partial _{x}\varphi _{\gamma
^{\prime }}\left( -x\right) \rangle $ with $\gamma ,\gamma ^{\prime }\in
\{\alpha ,\beta \}$, respectively. We see $\Omega _{\gamma \gamma ^{\prime
}} $ and $g_{\gamma \gamma ^{\prime }}^{y}$ are counteracting. Starting from
$\lambda =1$ we know from the QRM side $\varphi _{\alpha }$ and $\varphi
_{\beta }$ can be approximated by a displaced harmonic oscillator, while 
$\alpha ,\beta $ are positive to gain more negative
tunneling energy\cite{Ying2015}. Thus there is no zero point
before $g_{\mathrm{T1}}$. After $g_{\mathrm{T1}}$ there is one zero point in
$\psi _{\sigma_z }\left( x\right) $ so that $\psi _{\sigma_z}\left( x\right) $ is
braided once, which can be simply realized by changing the sign of $\alpha $ 
in $\psi _{-}$ (not the sign of $\beta$ otherwise the dominant same-side tunneling energy 
$\Omega _{\alpha \beta }$ becomes positive) and $\beta$ in $\psi _{+}$. Thus the main peak 
of $\psi _{-}$ flips (orange in Fig.\ref{fig-mechanism}(a,b)) and the parity is reversed.
Under this wave-function braiding, the same-side
energies $\Omega _{\alpha \beta }$ and $g_{\alpha \beta }^{y}$ are
unaffected and the dominant competition comes from the opposite-side ones $%
\Omega _{\gamma \gamma },g_{\gamma \gamma }^{y}$. The leading contributions
together are $E_{\Omega y}=\Omega _{\alpha \alpha }+g_{\alpha \alpha
}^{y}=\mp \alpha ^{2}(\frac{\Omega }{2}-\frac{2g_{y}g_{z}}{\omega })\exp
[-g_{z}^{\prime 2}]$ and the sign $\mp $ corresponds to before and after
the transition. Note the opposite-side kinetic and potential terms are not
competitive as their weight product $\alpha \beta $ from a same spin
component is much smaller than $\alpha ^{2}$ in the large $g$. Setting $%
E_{\Omega y}=0$ gives the transition boundary%
\begin{equation}
g_{\mathrm{T1}}=2g_{\mathrm{s}}/\sqrt{1-\lambda ^{2}},\quad \lambda _{%
\mathrm{T1}}=\sqrt{1-4g_{\mathrm{s}}/g^{2}}
\end{equation}%
which is frequency-independent and matches well with the numerics as in Fig.%
\ref{fig-PhaseDiagrams}. Further topological transitions similarly involve
more wave-function braidings, each time occurs with the energy reversing,
thus always associated with gap closing.

\textit{Conclusions.}--By mapping to a RSOC model we have
studied the phase transitions in the anisotropic QRM which connects the
fundamental QRM and JCM. By turning on finite frequencies we have unveiled a
novel phase diagram with a fine structure of phase boundary series in gap
closing and reopening, which is beyond the conventional picture in the low
frequency limit. The conventional transition is accompanied with a hidden
symmetry breaking, whereas the emerging gap-closing transitions are
topological transitions without symmetry breaking. We find the number of
zero points in the wave function is the topological number. We see this topological feature provides a 
novel universality classification in bridging the QRM and the JCM, among the diversity arising 
from going away from the low frequency limit where there is a universality\cite{LiuM2017PRL}. 
We have also shown that the conventionally 
established triple point is actually a quintuple or sextuple point and following the
penta-/hexa-criticality emerge a series of tetra-criticalities. Our work
demonstrates that a single-qubit system can exhibit multicriticalities and
topological transitions.


\textbf{Acknowledgements.}\textit{--}This work was supported by the National
Natural Science Foundation of China (Grant No. 11974151).

\end{document}